\documentstyle[12pt,epsfig]{article}
\author{Juli\'an Candia$^{a}$ and Ezequiel V. Albano$^{b}$\\{}\\
$^a${\small\it Departamento de F\'{\i}sica, UNLP, 
CC67, 1900 La Plata, Argentina}\\
$^b${\small\it Instituto de Investigaciones Fisicoqu\'{\i}micas
Te\'{o}ricas y Aplicadas}\\{\small\it (INIFTA), UNLP, CONICET, 
Suc.4, CC16, 1900 La Plata, Argentina}}

\title{Quasi-wetting and morphological phase transitions in  
confined far-from-equilibrium magnetic thin films}

\begin{document}
\maketitle

\begin{abstract}
The growth of confined magnetic films with ferromagnetic interactions between
nearest-neighbor spins is studied in a stripped $(1+1)-$dimensional rectangular
geometry. Magnetic films are grown irreversibly
by adding spins at the boundaries of the growing interface.
A competing situation with two opposite short range   
surface magnetic fields of the same magnitude 
is analyzed. Due to the antisymmetric condition considered, an interface
between domains with spins having opposite orientations develops 
along the growing direction. Such interface undergoes a 
localization-delocalization transition that is identified as  
a quasi-wetting transition, in qualitative
agreement with observations performed under equilibrium conditions. 
In addition, the film also exhibits a growing  
interface that undergoes morphological transitions 
in the growth mode. It is shown that, 
as a consequence of the nonequilibrium nature of 
the investigated model, the subtle interplay 
between finite-size effects, wetting, and interface 
growth mechanisms leads to more rich and complex
physical features than in the equilibrium counterpart.  
Indeed, a phase diagram that exhibits eight distinct regions 
is evaluated and discussed. In the thermodynamic limit, 
the whole ordered phase (which contains the 
quasi-wetting transition) collapses, while
within the disordered phase, standard extrapolation procedures show that 
only two regions are present 
in the phase diagram of the infinite system.     
\end{abstract}

\section{Introduction}

The preparation and characterization of magnetic nanowires and
films is of great interest for the development of advanced
microelectronic devices. Therefore, the study of the
behavior of magnetic materials in confined geometries,
e.g. thin films, has attracted both experimental
\cite{iron,tsay,w110} and theoretical
\cite{alba1,kar,ou,reis} attention.
On the other hand, the investigation of 
very interesting wetting phenomena has also drawn enormous attention.  
For instance, surface enrichment or wetting layers have been
observed experimentally in a great variety of systems, such as 
e.g. polymer mixtures \cite{poly1,poly3} and 
adsorption of simple gases on alkali metal surfaces \cite{alk3}.
Indeed, it is recognized that wetting of solid surfaces by a
fluid is a phenomenon of primary importance in many fields
of practical technological applications (lubrication, efficiency 
of detergents, oil recovery in porous material, stability of 
paint coatings, interaction of macromolecules with interfaces, etc.
\cite{degene}). 
Furthermore, 
the study of wetting transitions at interfaces has also attracted
considerable theoretical interest \cite{diet,for,parry},
involving, among others, different approaches such as the mean 
field Ginzburg-Landau method \cite{parr,swi}, transfer matrix 
and Pfaffian techniques \cite{macio1,macio2}, density matrix 
renormalization group methods \cite{car}, solving the
Cahn-Hilliard equation \cite{ch}, using Molecular Dynamic 
simulations \cite{md}, solving self-consistent field equations
\cite{scft}, and by means of extensive Monte Carlo 
simulations \cite{alba1,bind1,kur,mamu,kur1}.  

However, most of the theoretical work has been carried out
within the framework of equilibrium systems. In contrast,
the aim of this work is to study the properties of thin magnetic
film growth under far-from-equilibrium conditions, 
using extensive Monte Carlo simulations.
In order to simulate thin film growth,
our study is carried out in confined (stripped) geometries, which resemble
recent experiments where the growth of
quasi-one-dimensional strips of Fe on a Cu(111) vicinal surface
\cite{iron} and Fe on a W(110) stepped substratum \cite{w110}
have been performed.
Also, in a related context, the
study of the growth of metallic multilayers have shown a rich
variety of new physical phenomena. Particularly, the growth
of magnetic layers of Ni and Co separated by a Cu spacer
layer has recently been studied \cite{cobre}.
It should also be remarked that, although the 
discussion is presented here in terms of
a magnetic language, the relevant physical concepts can 
be extended to other systems such as fluids, polymers, and binary mixtures.  

In the present work it is shown that,
in far-from-equilibrium systems, the subtle interplay 
between finite-size effects, wetting, and interface 
growth mechanisms leads to more rich and complex
physical features than in the equilibrium counterpart.  
In fact, a complex phase diagram, that exhibits a 
localization-delocalization 
transition in the interface that runs along the walls and a 
change of the curvature of the
growing interface running perpendicularly to the walls, is
evaluated and discussed.

This manuscript is organized as follows: 
in Section II we give details on the simulation method,
Section III is devoted to the presentation and 
discussion of the results, 
while the conclusions are finally stated in Section IV. 

\section{The model and the simulation method}

In the classical Eden model \cite{eden} on the square lattice,
the growth process starts by adding particles to the
immediate neighborhood (the perimeter) of a seed particle.
Subsequently, particles are stuck at random to perimeter sites. 
This growth process leads to the formation of compact clusters with a 
self-affine interface \cite{shl1,shl2,bar,mar}.
The growth of a ferromagnetic material can be studied by means of the so called 
magnetic Eden model (MEM) \cite{mem}, which considers an additional
degree of freedom due to the spin of the growing particles.
In the present work the MEM is investigated on the square
lattice using a rectangular 
geometry $L\times M$ (with $M \gg L$ \cite{note}).

\begin{figure}
\centerline{{\epsfysize=2.2in \epsffile{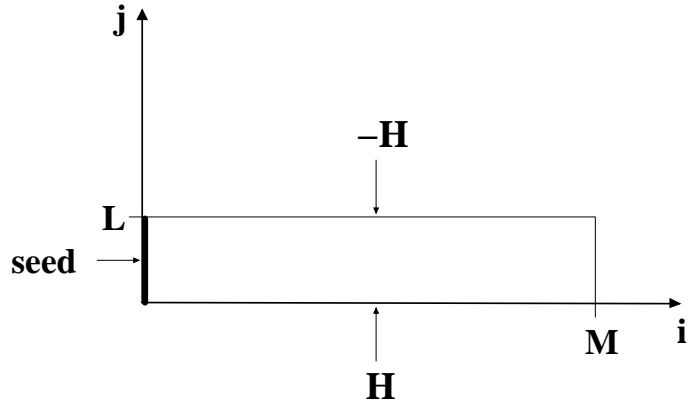}}}
\caption{The general set-up for the MEM in a $(1+1)-$dimensional rectangular 
geometry. The magnetic film grows along
the positive longitudinal direction from a seed constituted by
$L$ parallel-oriented spins placed at $i=1$,
as indicated. Open boundary conditions are assumed along the 
transverse direction, in which competing surface magnetic fields 
$H > 0$ ($H'=-H$) acting on
the sites placed at $j=1$ ($j=L$) are considered.
Since all deposited spins are frozen, an algorithm that shifts the active
growing region towards low $i$ values can be repeatedly applied 
when the film is
close to reaching the limit of the sample ($i=M$). Hence, finite-size results
are independent of $M$, and are thus only governed by the 
lattice width $L$.}
\label{FIG. 1}
\end{figure}

Figure 1 illustrates the general set-up assumed.
The location of each site on the
lattice is specified through its rectangular coordinates $(i,j)$,
($1 \leq i \leq M$, $1 \leq j \leq L$).
The starting seed for the growing cluster is a column
of parallel-oriented spins placed at $i=1$
and film growth takes place
along the positive longitudinal direction (i.e. $i \geq  2$).
The boundary conditions are open along the transverse direction,
in which competing surface magnetic fields $H > 0$ ($H'=-H$) acting on
the sites placed at $j=1$ ($j=L$) are considered. 
Then, magnetic films are grown by selectively adding
spins ($S_{ij}= \pm 1$) to perimeter sites, which are defined as the
nearest-neighbor (NN) empty sites of the already occupied ones.

Considering a ferromagnetic interaction of strength $J > 0$ 
between NN spins, the energy $E$ of a given configuration of 
spins is given by
\begin{equation}
E = - \frac{J}{2} \left( \sum_
{\langle ij,i^{'}j^{'} \rangle} S_{ij}S_{i^{'}j^{'}} \right) - 
 H  \left( \sum_{\langle i, \Sigma_1 \rangle } S_{i1} -
\sum_{\langle i, \Sigma_L \rangle } S_{iL}  \right)  \ \ , 
\end{equation}
\noindent where the summation $\langle ij,i^{ '}j^{ '}\rangle$
is taken over occupied NN sites, 
while $\langle i, \Sigma_1 \rangle$,
$\langle i, \Sigma_L \rangle$ 
denote summations carried over occupied sites on 
the surfaces $j=1$ and $j=L$, respectively.
Throughout this work we set the Boltzmann constant equal 
to unity and we take the temperature,
energy, and magnetic fields measured in units of $J$.
The probability for a perimeter site
to be occupied by a spin is taken to be
proportional to the Boltzmann factor 
$\exp(- \Delta E/T)$, where $\Delta E$
is the change of energy involved in the addition of 
the given spin. 
At each step, the probabilities of adding up and down
spins to a given site have to be evaluated
for all perimeter sites.
After proper normalization of the
probabilities, the growing site and the orientation of the spin
are determined through standard Monte Carlo techniques.
Although both the interaction energy and the
Boltzmann probability distribution considered for the MEM are
similar to those used for the Ising model 
with surface magnetic fields \cite{alba1},
it must be stressed that these two models
operate under extremely different conditions, namely the MEM 
describes the {\bf irreversible growth} of a magnetic material 
and the Ising model deals with a magnet under {\bf equilibrium}.
In the MEM, the position and orientation of all deposited
spins remain fixed.
During the growth process, the system develops a rough growth interface
and evolves mainly along the longitudinal direction. Some lattice sites 
can remain empty even well within the system's bulk, but, since at each 
growth step all perimeter sites are candidates for becoming occupied,
these holes are gradually filled. 
Hence, far behind the active growth
interface, the system is compact and frozen.    
When the growing cluster interface is
close to reaching the limit of the sample ($i=M$), 
the relevant properties of the irreversibly frozen
cluster's bulk (in the region where the growing process has definitively stopped)
are computed, the useless frozen bulk is thereafter erased, and
finally the growing interface is shifted 
towards the lowest possible longitudinal coordinate.
Hence, repeatedly applying this procedure, 
the growth process is not limited by the lattice length $M$,
and so the lattice width $L$ is the only relevant parameter
concerning the finite-size nature of the sample \cite{note}.
In the present work clusters having up to $10^{8}$ 
spins have typically been grown.

\section{Results and discussion}

Magnetic Eden films that grow in a confined geometry with competing
surface fields exhibit a very rich phase diagram, which is composed 
of eight regions (as shown in figure 2). These regions are delimited by several distinct,
well defined transition curves. As will be shown below, the bulk order-disorder
(finite-size) critical point $T_c(L)$, the Ising-like quasi-wetting transition
curve $T_w(L,H)$, and two morphological transitions associated to the 
curvature of the growing interface 
(namely, from convex to non-defined to concave), 
can be quantitatively located. Moreover, in order to gain some 
insight into the physics involved in this complex phase diagram, 
some typical snapshot configurations characteristic 
of the various different growth regimes are obtained (see figure 6) 
and discussed. 

\begin{figure}
\centerline{{\epsfysize=2.8in \epsffile{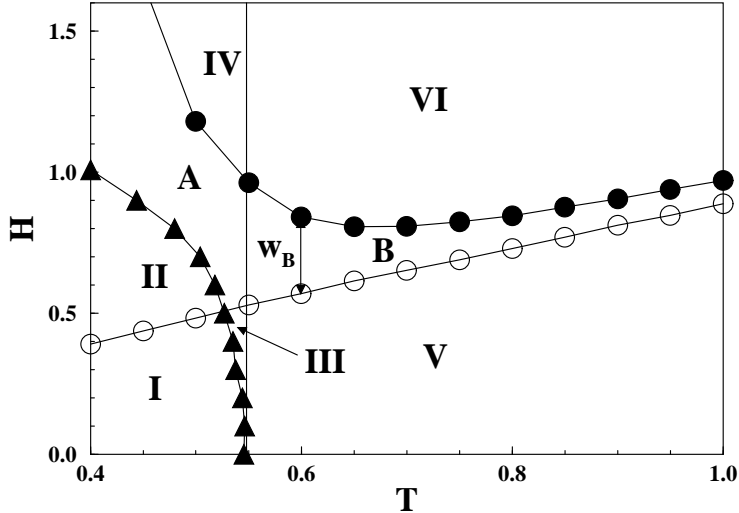}}}
\caption{$H-T$ phase diagram corresponding to a lattice of size $L=32$. 
The vertical straight line at $T_{c}(L) = 0.55$ corresponds 
to the $L-$dependent critical temperature, which
separates the low-temperature ordered phase
from the high-temperature disordered phase.
Open (filled) circles refer to the transition between non-defined and 
convex (concave) growth regimes, and triangles stand for the Ising-like 
localization-delocalization transition curve.
Eight different regions are distinguished, 
as indicated in the figure. Also $w_B$, the isothermal width of Region $B$, is 
marked for $T=0.6$. More details in the text.}
\label{FIG. 2}
\end{figure}

As well known from finite-size scaling theory, 
there is some degree of arbitrariness in locating  
the $L-$dependent critical temperature $T_c(L)$ of a
finite system. However, the critical point $T_c$ of the
infinite system, obtained by extrapolating $T_c(L)$ to the $L\rightarrow \infty$ limit,
is unique and independent of any particular choice for the finite-size critical point.
In particular, the ($L-$dependent) bulk order-disorder critical temperature can be 
identified with the peak of the susceptibility at zero surface field. 
For $L=32$, the critical point so defined is $T_{c}(L=32) = 0.55$,
and is shown in figure 2 by a vertical straight line.
So, the left (right) hand side part of
the phase diagram corresponds to the ordered (disordered) growth
regime that involves Regions $I,II,III,IV$, and $A$ 
(Regions $V,VI$, and $B$). 

Previous studies \cite{pre} have demonstrated that the MEM in a stripped 
$(1 + 1)-$dimensional geometry is not critical 
(i.e. it only exhibits an ordered phase at $T=0$ in the thermodynamic
limit). Hence, as we consider larger and larger lattices, 
the finite-size critical points $T_c(L)$ turn smaller and smaller, and
vanish indeed in the thermodynamic limit. This implies that
the eight regions coexisting in the $H-T$ phase diagram are
a finite-size effect only relevant for the growth of magnetic films in confined
geometries. The ordered phase corresponding to
the $T<T_c(L)$ region becomes steadily narrower the larger the lattice width $L$,   
and in the $L\rightarrow \infty$ limit only the disordered phase corresponding to 
the $T>T_c(L)$ region survives. Furthermore, as will be shown below, also  
Region $B$ shrinks and collapses in this limit, so that only Regions $V$ and $VI$ are present
in the phase diagram of the infinite system. 

As in previous investigations \cite{pre}, 
let us define the mean transverse magnetization $m(i,L,T,H)$ as
\begin{equation}
m \ \ (i,L,T,H) = \frac{1}{L} \sum_{j = 1}^{L} S_{ij} \ \ .      
\end{equation}
\noindent Also the susceptibility $\chi$
can be defined, as usual, in terms of the magnetization fluctuations.
Then, using a standard procedure \cite{alba1}, 
the  localization-delocalization transition  
curve corresponding to the up-down interface running along the walls
can be computed considering that on the $H-T$ plane,
a point with coordinates $(H_w,T_w)$ on this curve
maximizes $\chi(H,T)$. 
So, the size-dependent localization-delocalization transition curve
is obtained, as shown in figure 2 (curve with triangles).
As in the case of the Ising model, this quasi-wetting 
transition refers to a transition
between a nonwet state that corresponds to a localized 
interface bound to one of the confinement walls, 
and a wet state associated to a 
delocalized domain interface centered between roughly equal domains of
up and down spins \cite{alba1,swi}.
In fact, it is observed a finite jump in the wetting layer thickness
that takes place as a result of the finite size of the system.
As the lattice size is increased, the magnitude of the jump grows.
However, it should be remarked again that the occurrence of this phenomenon at
finite temperature is essentially due to the small size of the thin film,
and it becomes irrelevant in the thermodynamic limit. Since confined $(1+1)-$
dimensional magnetic Eden films are noncritical \cite{pre},
the whole ordered phase corresponding to $T<T_c(L)$ 
(involving Regions $I,II,III,IV$, and $A$) 
that also contains the quasi-wetting curve, 
collapses in the $L\rightarrow \infty$ limit. Thus, standard procedures  
carried out in the investigation of equilibrium wetting phenomena 
concerning the finite-size scaling behavior within the ordered phase 
\cite{alba1,parr,swi,macio1,macio2,car,ch,md,scft,bind1,kur,mamu,kur1}
are simply meaningless in the case of the model investigated here. 

Since the MEM is a nonequilibrium kinetic growth model, 
it also allows the identification of another kind of phase transition, 
namely a morphological transition associated with the curvature of the growing 
interface of the system. To avoid confusion, we 
remark that the term {\it interface} is used here for the 
transverse interface between occupied and empty lattice sites, 
while it was used above for the longitudinal
interface between up and down spin domains.

\begin{figure}
\centerline{{\epsfysize=3.2in \epsffile{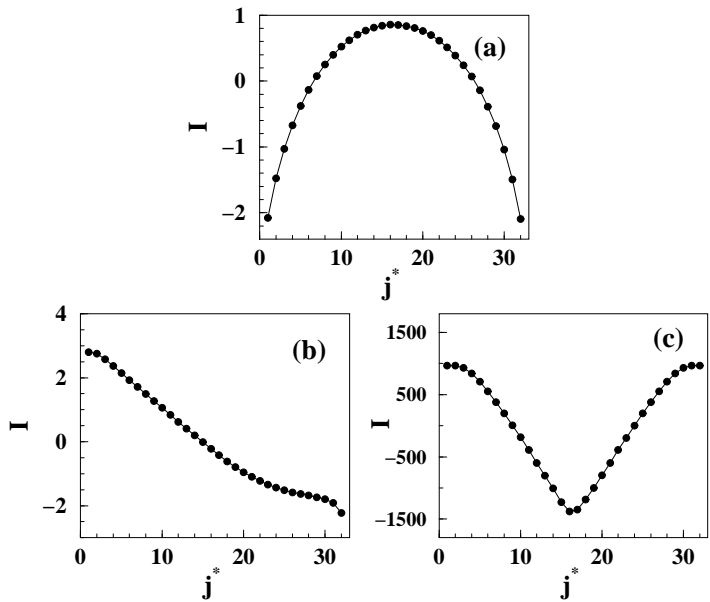}}}
\caption{Plots of the averaged interface profile 
$I$ vs $j^{*}$ for $T=0.6$, $L=32$ and 
different values of the surface magnetic field $H$:
(a) $H=0$, (b) $H=0.6$, and (c) $H=4$.
The side $j^{*}=1$ ($j^{*}=L$) is the one associated with 
the dominant (non-dominant) spin domain. 
Increasing the surface fields, the curvature of the growing interface changes:
convex (a) $\rightarrow$ non-defined (b) $\rightarrow$ concave (c). 
This qualitative behavior has been observed for all temperatures and 
lattice sizes within the range of interest of this work.}
\label{FIG. 3}
\end{figure}

Figure 3 shows the shape of the mean growing interface $I$ obtained 
for different values of the surface magnetic field $H$, for fixed
temperature ($T=0.6$) and lattice size ($L=32$). Notice that the transverse coordinate
has been conveniently redefined, so that $j^{*}=1$ ($j^{*}=L$) corresponds to    
the side of dominant (non-dominant) spin domain, while
$j=1$ ($j=L$) is the side of positive (negative) magnetic field.       
>From the figure it follows that three qualitatively distinct growth 
regimes can clearly be distinguished. Indeed, it is observed that, 
while for small fields the system grows with convex curvature, 
increasing the fields the growth process enters into a regime of
non-defined curvature, since the dominant spin domain partially 
wets the confinement wall, while the non-dominant domain does not.
But then, further increasing the fields, a point is reached 
where the non-dominant spin domain also (partially) wets the wall and the growing 
interface turns concave. This qualitative
behavior has been observed for all temperatures and lattice 
sizes within the range of interest of this work. 
 
\begin{figure}
\centerline{{\epsfysize=2.5in \epsffile{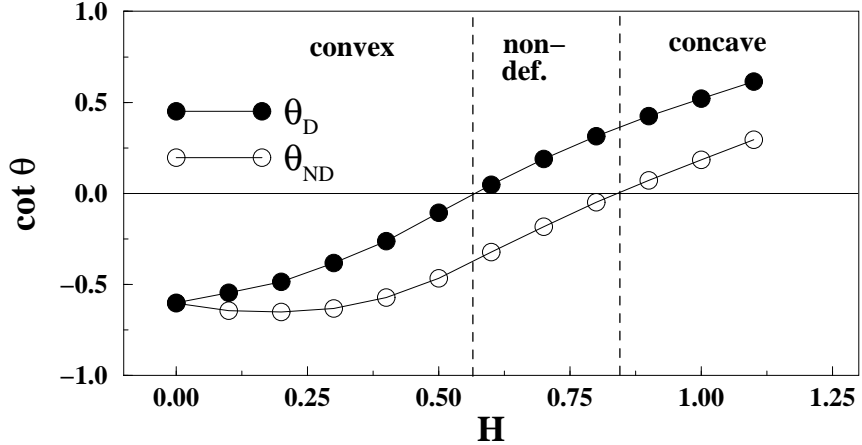}}}
\caption{Plots of $\cot(\theta)$ vs $H$ for $T=0.6$ and $L=32$.
$\theta_D$ ($\theta_{ND}$) is the contact angle corresponding to 
the dominant (non-dominant) spin cluster, and is represented 
by open (filled) circles.
The vertical dashed lines mark the fields that separate a given growth 
regime from another one, as indicated. A reference line corresponding to
$\cot(\theta)=0$ has also been included.}
\label{FIG. 4}
\end{figure}

To explore this phenomenon quantitatively, the behavior
of the contact angles between the growth interface and 
the confinement walls (as functions of temperature and magnetic field) have
to be investigated. Clearly, two different contact
angles should be defined in order to locate this transition,
namely $\theta_D$ for the angle corresponding to the 
dominant spin cluster, and $\theta_{ND}$ for the one that 
corresponds to the non-dominant spin cluster.
Figure 4 shows plots of $\cot(\theta)$ vs $H$ 
for $T=0.6$ and $L=32$. The vertical dashed
lines indicate the fields that separate a given growth regime from another one.
One observes that, increasing the surface fields, the growth regime 
changes from convex to non-defined to concave, in agreement with the 
interface profiles plotted in figure 3.   
Analogously, figures 5(a)-(d) show plots of $\cot(\theta)$ vs $T$ 
for $L=32$ and several different values for the magnetic field $H$.
Again, vertical dashed lines correspond to transition temperatures 
between different growth regimes. Figure 5(a) corresponds to $H=0.2$ 
and displays the characteristic behavior for 
very small magnetic fields, that is, a convex growing interface 
irrespective of temperature.  
\begin{figure}
\centerline{{\epsfxsize=3.3in \epsffile{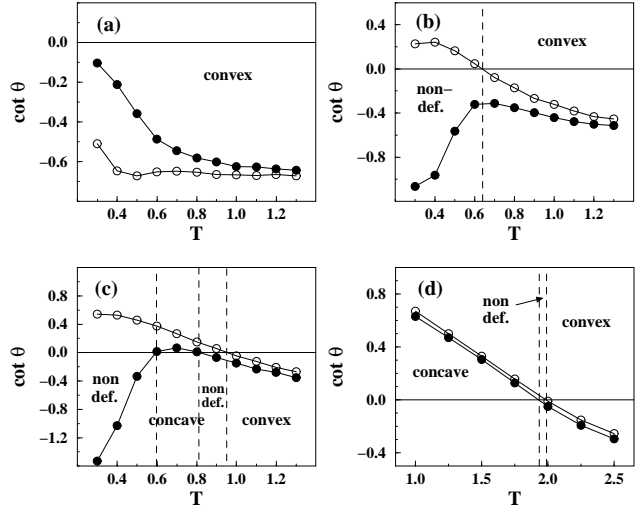}}}
\caption{Plots of $\cot(\theta)$ vs $T$ for $L=32$
and several different magnetic fields: 
(a)$H=0.2$, (b)$H=0.6$, (c)$H=0.85$, and (d)$H=1.8$.  
$\theta_D$ ($\theta_{ND}$) is the contact angle corresponding to 
the dominant (non-dominant) spin cluster, and is represented 
by open (filled) circles.
The vertical dashed lines mark the temperatures that separate a given growth 
regime from another one, as indicated. Reference lines corresponding to
$\cot(\theta)=0$ have also been included.}
\label{FIG. 5}
\end{figure}

For $H=0.6$ one observes a single transition from the growth 
regime of non-defined curvature to the convex growth regime, 
which shows up by increasing the temperature, 
as shown in figure 5(b). It should be noticed that the concave 
growth regime is prevented, since for small enough magnetic
fields $\cot(\theta_{ND}) < 0$ for all $T$. As the fields are increased,
$\cot(\theta_{ND})$ moves upwards and crosses 
$\cot(\theta_{ND})=0$, as expected from the plot of figure 4. 
For instance, the plots of $\cot(\theta)$ vs $T$ 
for $H=0.85$, shown in figure 5(c), exhibit this behavior. Hence, here one
has to deal with three transition temperatures. 
Finally, by further increasing the fields,
the whole low-temperature region is dominated by the concave growth 
regime and two transition temperatures remain, 
as shown in figure 5(d) for $H=1.8$.  
All these features are compactly shown in the $H-T$ phase diagram 
of figure 2, where open (filled) circles refer to the transition 
between non-defined and  convex (concave) growth regimes. 

As above anticipated, we will now introduce and discuss 
some characteristic snapshot pictures, 
in order to provide qualitative explanations that account for 
the different growth regimes observed.
Let us begin with Region $I$ (see figure 2), that corresponds 
to the Ising-like nonwet state and the convex growth regime. 
In this region, the temperature is small and the
system grows in an ordered state, i.e. the dominant spin domain prevails and
the deposited particles tend to have their spins all pointing in 
the same direction. Small clusters with the opposite orientation 
may appear preferably on the surface where the non-dominant 
orientation field is applied. These ``drops'' might 
grow and drive a magnetization reversal, thus changing the sign 
of the dominant domain (see figure 6(a)). 
In fact, the formation of sequences of 
well-ordered domains are characteristic of the ordered phase of 
confined (finite-size) spin systems
such as the Ising magnet \cite{alba1}. 
Due to the open boundary conditions, 
perimeter sites at the confinement walls 
experience a missing neighbor effect, that is, the number
of NN sites is lower than for the case of perimeter sites
on the bulk. 
\begin{figure}
\centerline{{\epsfysize=2.8in \epsffile{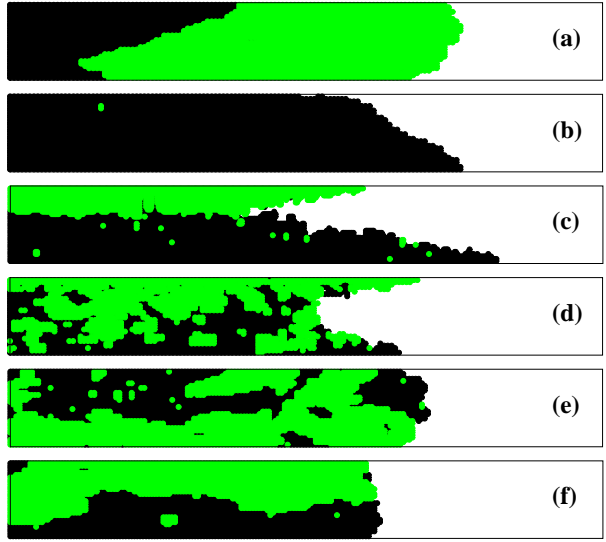}}}
\caption{Typical snapshot configurations that exhibit
a variety of different growth regimes. 
Gray (black) circles correspond to spins up (down). 
The surface field on the upper (lower) confinement
wall is positive (negative). 
The snapshots correspond to a lattice size $L=32$ and 
several different values of temperature and surface fields: 
(a)$H=0.05$, $T=0.4$; (b)$H=0.7$, $T=0.4$;
(c)$H=1.8$, $T=0.5$; (d)$H=1.8$, $T=1.0$; (e)$H=0.1$, $T=0.7$;
and (f) $H = 0.3$, $T = 0.54$.}
\label{FIG. 6}
\end{figure}
Since $H$ is too weak to compensate this effect, 
the system grows preferentially along the center of the sample 
as compared to the walls, and the
resulting growth interface exhibits a convex shape.
So, Region $I$ corresponds to the Ising-like nonwet
state and the convex growth regime, as shown by the snapshot picture
of figure 6(a).

Let us now consider an increase in the fields, 
such that we enter Region $II$ (see figure 2).
Since the temperature is kept low, the system is still in its ordered phase and
neighboring spins grow preferably parallel-oriented. The surface fields
in this region are stronger and thus capable of compensating the missing
NN sites on the surfaces. But, since the fields on both surfaces have opposite
signs, it is found that, on the one hand, the field that 
has the same orientation as the dominant spin cluster favors the 
growth of surface spins, while on the other hand,
the sites on the surface with opposite field have a lower probability to
be chosen during the Monte Carlo growth process. Hence, 
the contact angle corresponding to
the dominant spin cluster is then $\theta_D < \frac{\pi}{2} $, while the
non-dominant is $\theta_{ND} > \frac{\pi}{2} $. 
Thus, on the disfavored side the growing interface becomes pinned
and the curvature of the growing interface is not defined. 
Figure 6(b) shows a typical snapshot corresponding to Region $II$.

Keeping $H$ fixed within Region $II$ but increasing the temperature, 
thermal noise will enable the formation of drops on the disfavored side 
that eventually may nucleate into larger clusters as the temperature 
is increased even further. This process may lead to the emergence of 
an up-down interface, separating oppositely oriented domains, running in the
direction parallel to the walls. Since sites along
the up-down interface are surrounded by oppositely oriented NN spins, 
they have a low growing probability. 
So, in this case the system grows preferably along the confinement 
walls and the growing interface is concave (figure 6(c)). 
Then, as the temperature is
increased, the system crosses to Region $A$ (see figure 2) and we observe 
the onset of two competitive growth regimes: {\it (i)} one  
exhibiting a non-defined growing curvature that appears when 
a dominant spin orientation 
is present, as in the case shown in figure 6(b); 
{\it (ii)} another that appears
when an up-down interface is established and the system
has a concave growth interface, as is shown in figure 6(c). 

Further increasing the temperature and for large enough fields, 
the formation of a stable longitudinal up-down interface
that pushes back the growing interface is observed. So,
the system adopts the concave growth regime (see figure 6(c) corresponding to
Region $IV$ in figure 2). 
Increasing the temperature beyond $T_c(L)$,
a transition from a low-temperature ordered 
state (Region $IV$) to a high-temperature disordered state
(Region $VI$, see figure 6(d)), both within the concave 
growth regime, is observed. Analogously, for small enough fields, 
a temperature increase drives the system from the ordered convex growth
regime (Region $I$) to the disordered convex growth regime (Region $V$,
see figure 6(e)). As shown in figure 2, there is also an 
intermediate fluctuating state
(Region $B$) between Regions $V$ and $VI$, 
characterized by the competition between
the disordered convex growth regime and the disordered concave one.  

Finally, a quite unstable and small region (Region $III$ in figure 2) that
exhibits the interplay among the growth regimes of the contiguous
regions, can also be identified. Since the width of Region $III$
is of the order of the rounding observed in $T_c(L)$, large
fluctuations between ordered and disordered states are observed,
as well as from growth regimes of non-defined curvature to  
convex ones. However, figure 6(f) shows a snapshot configuration 
that is the fingerprint of Region $III$, 
namely a well defined spin up-down interface with an almost
flat growing interface. 

\begin{figure}
\centerline{{\epsfysize=2.5in \epsffile{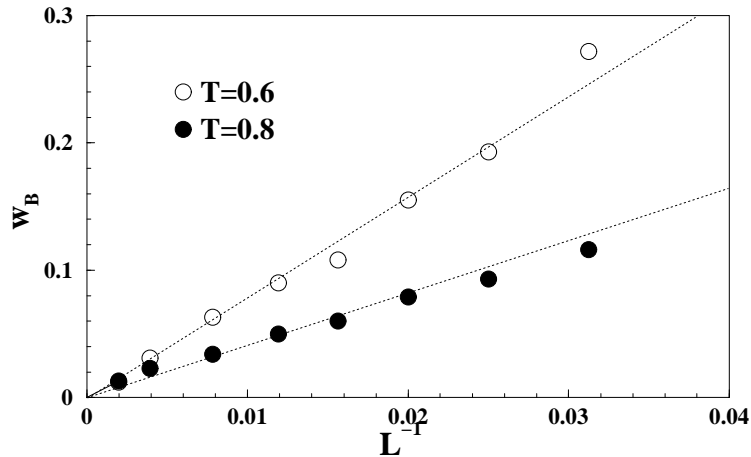}}}
\caption{Plots of the isothermal width of Region $B$ ($w_B(L,T)$) 
versus $L^{-1}$ for lattice sizes in the range $32\leq L\leq 512$, 
corresponding to $T=0.6$ and $T=0.8$. The lines are guides to the eye.
Following this extrapolation procedure, it turns out that Region $B$
collapses in the $L\to\infty$ limit, and so only Regions
$V$ and $VI$ are present in the infinite size phase diagram.}
\label{FIG. 7}
\end{figure}

As already commented, the noncriticality of confined $(1+1)-$
dimensional magnetic Eden films implies that 
the whole ordered phase corresponding to the $T<T_c(L)$ region, 
which contains the quasi-wetting curve, 
collapses in the $L\rightarrow \infty$ limit, since in this limit 
$T_c(L)\rightarrow 0$ \cite{pre}. 
Concerning the structure of the disordered phase in the thermodynamic limit,
it can be shown that the morphological transition curves merge into a single
curve that separates the disordered convex growth regime (Region $V$) from
the disordered concave one (Region $VI$). 
In order to illustrate this, we define $w_B(L,T)$ as the isothermal width of
the intermediate fluctuating state (Region $B$). For instance, $w_B$ is marked 
in figure 2 for $L=32$ and $T=0.6$. 
Figure 7 shows plots of $w_B(L,T)$ versus $L^{-1}$ corresponding to 
lattice sizes in the range $32\leq L\leq 512$ and 
for two different temperatures.  
Using this standard extrapolation procedure, it turns out that 
$w_B(L,T)\to 0$ as $L\to\infty$, and so only Regions
$V$ and $VI$ are present in the infinite size phase diagram.

\section{Conclusions}

The growth of magnetic Eden thin films with
ferromagnetic interactions between nearest-neighbor spins has
been studied in a $(1+1)-$dimensional geometry with competing 
surface magnetic fields.
Besides an Ising-like quasi-wetting transition, two
morphological transitions in the growing interface, which arise
from the MEM's kinetic growth process, have also been identified.
The resulting phase diagram exhibits eight regions. 
Among them, six correspond to well defined growth regimes, 
which are illustrated by typical snapshot pictures.
The remaining two regions appear as
fluctuating crossover states characterized by the competition between
the growth regimes of neighboring regions.   
We hence conclude that the nonequilibrium nature of growing magnetic
Eden thin films introduces new and rich physical features of interest,
as compared to wetting phase diagrams for equilibrium spin systems 
(e.g. the well-studied Ising model \cite{alba1}). 
In the thermodynamic limit, the whole ordered phase (which contains the 
quasi-wetting curve) collapses. Indeed only two regions, 
which correspond to the disordered convex growth regime and
the disordered concave one, are present in the phase diagram of
the infinite system.   

We expect that the present study will contribute to 
the fields of nonequilibrium wetting phenomena 
and irreversible growth processes in confined geometries, 
and we hope that it will stimulate further experimental and theoretical 
work in these topics of widespread technological and scientific interest.  

\section*{Acknowledgments} 
This work was supported by CONICET, UNLP, and ANPCyT (Argentina).


\begin{thebibliography}{99}
\bibitem{note} Notice that $M$ is the growing direction of the
magnetic film that plays the role of the growing ``time''. After 
reaching the stationary growing regime this variable is irrelevant 
for the computation of averaged quantities and consequently does not 
enter in the scaling functions.
\end{thebibliography}
\end{document}